\documentclass[10pt,journal]{IEEEtran}
\usepackage{amsmath,amsfonts}
\usepackage{algorithmic}
\usepackage{array}
\usepackage[caption=false,font=normalsize,labelfont=sf,textfont=sf]{subfig}
\usepackage{textcomp}
\usepackage{stfloats}
\usepackage{url}
\usepackage{verbatim}
\usepackage{graphicx}
\def\BibTeX{{\rm B\kern-.05em{\sc i\kern-.025em b}\kern-.08em
    T\kern-.1667em\lower.7ex\hbox{E}\kern-.125emX}}
\usepackage{balance}
\usepackage{booktabs}
\usepackage{etoolbox}
\makeatletter\patchcmd{\@makecaption}{\scshape}{}{}{}
\usepackage{color}
\usepackage{amsthm,amssymb,amsfonts}
\usepackage[linesnumbered,ruled,lined]{algorithm2e}
\usepackage{bm}
\usepackage{cases}
\newtheorem{proposition}{Proposition}

\newtheorem{lemma}{Lemma}

\usepackage[colorlinks,
linkcolor=blue,       
anchorcolor=blue,  
citecolor=blue,        
]{hyperref}

\newcommand{\E}{\mathbf E}

\newcommand{\w}{\mathbf w}

\begin{document}
\title{A Low-Complexity Beamforming Design for Beyond-Diagonal RIS aided Multi-User Networks}
\author{Tianyu Fang, Yijie Mao, \textit{Member, IEEE}
\thanks{This work has been supported in part by the National Nature Science Foundation of China under Grant 62201347; and in part by Shanghai Sailing Program under Grant 22YF1428400.
	\par T. Fang and Y. Mao are with the School of Information Science and Technology, ShanghaiTech University, Shanghai 201210, China (e-mail:
	{fangty, maoyj}@shanghaitech.edu.cn).}
\vspace{-0.8cm} 
 
 }


\maketitle

\begin{abstract}
Beyond-diagonal reconfigurable intelligent surface (BD-RIS) has been proposed recently as a novel and generalized RIS architecture that offers enhanced wave manipulation flexibility and large coverage expansion. However, the beyond-diagonal mathematical model in BD-RIS inevitably introduces additional optimization challenges in beamforming design.
In this letter, we derive a closed-form solution for the BD-RIS passive beamforming matrix that maximizes the sum of the effective channel gains among users. We further propose a computationally efficient two-stage beamforming framework to jointly design the active beamforming at the base station and passive beamforming at the BD-RIS to enhance the sum-rate for a BD-RIS aided multi-user multi-antenna network.
Numerical results show that our proposed algorithm achieves a higher sum-rate while requiring less computation time compared to state-of-the-art algorithms. The proposed algorithm paves the way for practical beamforming design in BD-RIS aided wireless networks.
\end{abstract}  


\vspace{-0.3cm}
\section{Introduction}
\par Reconfigurable intelligent surface (RIS) has emerged as a promising enabling technology for the sixth generation of wireless networks \cite{Wu2020,Lin2022}. 
In recent advancements, a groundbreaking architecture of RIS called beyond-diagonal reconfigurable intelligent surface (BD-RIS) has been proposed  and garnered recognition as a revolution within the realm of RIS \cite{Shen2022}. Unlike conventional RIS models, the interconnection between reflecting elements in BD-RIS is characterized by the scattering parameter matrix in microwave theory. By such means, BD-RIS can be further classified into the single, group, and fully connected architectures according to different performance and hardware complexity trade-off \cite{Li2023}. The fully connected architecture offers the highest level of flexibility and optimization potential but comes with increased complexity and control requirements. The single connected architecture, which has been widely investigated in existing works, provides a simpler implementation but has limited performance. The group connected architecture divides the RIS elements into groups, enabling performance improvements while maintaining manageable complexity.

\par The study of BD-RIS is still in its infancy. An optimal passive beamforming design has been proposed in \cite{Nerini2023a} to maximize the BD-RIS assisted channel gain only in point-to-point single-antenna networks, this approach, however, becomes inefficient when extending to multi-user multi-antenna transmission. Besides, existing works on BD-RIS \cite{Li2023,Li2023a,Fang2022} all address the joint optimization problem of active beamforming at the base station (BS) and the passive beamforming at the BD-RIS using the alternative optimization (AO) framework, but this approach suffers from slow convergence and extensive computation time, limiting its implementation within each coherence block.

In this work, we first propose a closed-form solution for BD-RIS passive beamforming, aiming to maximize the sum of the effective channel gains among users. Based on this solution, a two-stage beamforming design framework is then established to jointly design the passive beamforming at the BD-RIS and active beamforming at the BS for sum-rate maximization. We demonstrate that the proposed framework achieves higher sum-rate  while significantly reducing the computation time than the state-of-the-art AO framework. This substantial reduction in computational complexity makes our framework highly practical for  BD-RIS-aided  multi-user multi-antenna transmission networks, facilitating their real-world deployment.

\section{System Model and Problem Formulation}
\label{system}

\subsection{Transmission Model}

Consider a BD-RIS assisted multi-user multiple-input single-output (MU-MISO) communication network comprising a BS with $ L $ antennas, a BD-RIS with $ N $ passive reflecting elements, and a set of $ K $ single-antenna users indexed as $ \mathcal K=\{1,\cdots,K\} $. The BS serves all $ K $ users simultaneously with the assistance of one BD-RIS. The passive beamforming matrix (also known as the scattering matrix) of the BD-RIS is denoted as $ \bm \Theta \in\mathbb C^{N\times N}$. We respectively denote the channels between the BS and user $ k $ as $ \mathbf g_k\in\mathbb{C}^{L\times 1} $, between the BD-RIS and user $ k $ as $ \mathbf h_k\in\mathbb C^{N\times 1} $, and between the BS and BD-RIS as $ \mathbf E\in\mathbb C^{N\times L} $. Let $ \mathbf s= [s_1,\cdots,s_K]^T\in\mathbb C^{K\times 1} $ represent the data stream vector containing the data streams intended for the respective users, and $ \mathbf W=[\mathbf w_1,\cdots,\mathbf w_K]\in\mathbb C^{L\times K} $ be the corresponding active beamforming matrix at the BS. The transmitted signal at the BS can be expressed as $ \mathbf x=\sum_{k=1}^K \mathbf w_k s_k \in\mathbb{C}^{L\times 1} $. Assuming each stream $ s_k,k\in\mathcal K $ has zero mean and unit variance, i.e., $ \mathbb{E}[\mathbf s\mathbf s^H]=\mathbf I $, we have a transmit power constraint given by $ \|\mathbf W\|^2_F\leq P_t $, where $ \|\cdot\|_F $ refers to the Frobenius norm and $ P_t $ represents the maximum transmit power of the BS. The signal is transmitted through both the direct signal path from the BS to the users and the BD-RIS aided path. Thus, the receive signal at user $ k $ is:
\vspace{-0.15cm}
\begin{equation}\label{eq:received_signal}
	y_k=\left(\mathbf g^H_k+\mathbf h_k^H\bm\Theta\E\right)\sum_{i=1}^{K}\mathbf w_i s_i+n_k,
	\vspace{-0.15cm}
\end{equation}
where $ n_k\sim\mathcal{CN}(0,\sigma_k^2) $ represents the additive white Gaussian noise (AWGN) with zero mean and variance $ \sigma_k^2 $.
\vspace{-0.15cm}
\subsection{Beyond-Diagonal Reconfigurable Intelligent Surface model}
\vspace{-0.05cm}
In this study, we aim to enhance the system performance by employing a BD-RIS  \cite{Shen2022} to assist the transmission from the BS to the users. All three categories of BD-RIS, including the single, group, and fully connected BD-RISa are considered in this paper. 
\par The fully connected BD-RIS consists of a reconfigurable impedance network, where each port is interconnected with other ports through a reconfigurable reactance. Consequently, the scattering matrix $ \mathbf \Theta $ of the fully connected BD-RIS should satisfy the following constraint:
\[ \mathcal M_1=\left\{\bm\Theta \big| \bm\Theta=\bm\Theta^T, \bm\Theta\bm\Theta^H=\mathbf I\right\}.\]

\par 
When the reconfigurable impedance network disconnects each port from the others, the fully connected BD-RIS simplifies to a single connected RIS, thereby satisfying the following constraint:
\vspace{-0.1cm}
\[\mathcal M_2=\left\{\bm\Theta\big|\bm \Theta=\mathrm{diag}\left( e^{j\theta_1},e^{j\theta_2},\cdots, e^{j\theta_N}        \right)\right\},\]
where $ \theta_n\in[0,2\pi) $ denotes the phase shift angle.

Furthermore, for the group connected BD-RIS  proposed in \cite{Shen2022}, the $ N $ BD-RIS elements are divided into $ G $ groups, with each group containing $ N_g $ elements. Each element within a group connects to all other elements in the same group while disconnecting to elements in other groups. Consequently, the scattering matrix takes the form of a block diagonal matrix as:
\[\mathcal M_3\!=\!\left\{\!\bm\Theta\big|	\bm\Theta\!=\!\mathrm{diag}(\!\bm\Theta_1,\cdots,\bm\Theta_G), \!\bm \Theta_g^H\!\bm \Theta_g\!=\!\mathbf I,\!\mathbf \Theta_g\!=\!\mathbf \Theta_g^T,\forall g\right\},\]
where $ \bm\Theta_g, g\in\{1,\cdots,G\} $ are complex symmetric unitary matrices. 

\vspace{-0.2cm}
\subsection{Problem Formulation}
The first objective in this work is to design the passive beamforming matrix of BD-RIS to maximize the sum of effective channel gains among users, which is formulated as:
\vspace{-0.2cm}
\begin{subequations}\label{P2}
	\setlength\abovedisplayskip{-5pt}
		\begin{align}
			\max_{\bm\Theta}\,\, & \sum_{k=1}^K \|\mathbf g_k^H+\mathbf h_k^H\mathbf \Theta\mathbf E\|^2\\
			\label{unic}	\text{s.t.}\,\,	&\bm\Theta\in\mathcal M_i,\,\forall i\in\{1,2,3\},
			\vspace{-0.1cm}
		\end{align}
\end{subequations}
Problem \eqref{P2} is non-convex due to the non-concave objective function and the non-convex constraints $ \mathcal M_i,\forall i\in\{1,2,3\} $.

Note that a similar problem has been addressed in \cite{Nerini2023a}, specifically tailored for the wireless power transfer application. In contrast to the objective function in \cite{Nerini2023a}, which is based on the $ l2-$norm of the effective channel matrix, our work utilizes the sum channel gain as the objective function. This is equivalent to the square of Frobenius norm for the effective channel matrix. Therefore, there is a fundamental difference on the problems considered in \cite{Nerini2023a} and in this work for the considered MU-MISO network. Attempting to design a passive beamforming matrix using the algorithm proposed in \cite{Nerini2023a} results in undesirable performance for our specific problem \eqref{P2}.

\begin{flushright}
	
\end{flushright}
\vspace{-1cm}
 
\section{Low-Complexity Passive Beamforming Design}
\label{Section framework}
In this section, we propose a symmetric unitary projection method with reduced computational complexity for solving problem \eqref{P2}, which involves the following two steps:
\begin{itemize}
	\item In the first step, we relax the non-convex constraints onto a convex set and solve the relaxed problem.
	\item In the second step, the relaxed solution is projected onto the feasible point within the non-convex sets $ \mathcal M_i,\,\forall i\in\{1,2,3\} $.   
\end{itemize}
\vspace{-0.2cm}
\subsection{Relaxed Passive Beamforming Design}
To derive a tractable solution of the passive beamforming matrix, we first relax the non-convex sets $ \mathcal M_i,\forall i\in\{1,2,3\} $ onto a convex sphere set $ \mathcal S=\left\{\bm\Theta\big| \|\mathbf \Theta\|_F^2  \leq N \right\} $. Obviously, \ $ \mathcal M_i\subset \mathcal S , \forall i\in\{1,2,3\} $. The relaxed problem becomes:
\begin{subequations}\label{P3}
	\begin{align}
		\max_{\bm\Theta}\,\, &f(\bm\Theta)\triangleq\big\|\mathbf G^H+\mathbf H^H\bm\Theta\E \big\|_F^2\\
	\label{trace}	\text{s.t.}\,\,	&\bm\Theta\in\mathcal S,
	\end{align}
\end{subequations}
 Where $ \mathbf G\triangleq[\mathbf g_1,\cdots,\mathbf g_K] $ and $ \mathbf H\triangleq[\mathbf h_1,\cdots,\mathbf h_K] $. We note that the objective function $ f(\bm\Theta) $ is an equivalent transformation of the objective function in problem \eqref{P2}.

\par 
Though problem \eqref{P3} is non-convex, it can be optimally solved by vectorization and eigenvalue decomposition as shown in the \cite[Lemma 1]{Demir2022}. Here, we give out the optimal solution of problem \eqref{P3} in the following.
\par \textit{Optimal solution of \eqref{P3}}:
	The objective function $ f(\bm\Theta) $ can be further rewritten as $ \|\mathbf a+\mathbf A \mathrm{vec}(\bm\Theta)\|^2 $, where $\mathbf A\triangleq \mathbf E^T\otimes \mathbf H^H$, $ \mathbf a=\mathrm{vec} (\mathbf G^H) $, $ \otimes $ denotes the Kronecker product and $ \mathrm{vec}(\cdot) $ refers to the vectorization operation. Denote the eigenvalues and corresponding orthogonal eigenvectors of $ \mathbf A^H\mathbf A $ as $ \lambda_d $ and $ \mathbf q_d $ respectively, the optimal solution $ \bm\Theta^\star $ of \eqref{P3} is given by
	\begin{equation}\label{optimal}
			\setlength\abovedisplayskip{0pt}
		 \mathrm{vec}(\bm\Theta^\star)=\sum_{d=1}^{N^2}\frac{\mathbf q_d\mathbf q_d^H\mathbf A^H\mathbf a}{\gamma^\star-\lambda_d},
	\end{equation}
where $ \gamma^\star $ is the unique root of \begin{equation}\label{key}
		\setlength\abovedisplayskip{0pt}
	\sum_{d=1}^{N^2}\frac{\big|\mathbf q_d^H\mathbf A^H\mathbf a\big|^2}{\gamma-\lambda_d}=N.
\end{equation} 
 The detail proof for \eqref{optimal} follows the procedure in \cite{Demir2022} except vectorization. Obviously, the optimal solution \eqref{optimal} involves a high-demensional eigenvalue decomposition and a bisection search to find the optimal dual variable $ \gamma^\star $, leading to unaffordable computational complexity order of $ \mathcal{O}(I_{\gamma}N^6) $, where $ I_{\gamma} $ is the iteration number for the bisection search to find the optimal $ \gamma^\star $. Therefore, this solution is not practical.
\par \textit{Proposed low-complexity solution of \eqref{P3}}: To derive a practice solution while maintaining a nearly optimal performance, we propose the following solution based on the gradient decent approach at an initial point $ \bm\Theta_0=\mathbf 0 $. Specifically, for problem \eqref{P3}, the gradient of $ f(\bm\Theta) $ at $ \bm\Theta_0=\bm 0 $ is given by
\begin{equation}\label{grad}
	\nabla_{\bm\Theta} f(\bm\Theta_0)=\mathbf H\mathbf G^H\mathbf E^H. 	
\end{equation}
We directly take \eqref{grad} as the decent direction $ \mathbf D_0=\nabla_{\bm\Theta} f(\bm\Theta_0) $ and use Armijo rule to determine a step size $ \alpha $ such that
\begin{equation}\label{armijo}
	f(\bm\Theta_0+\alpha\mathbf D_0)\geq f(\bm\Theta_0) +\zeta\cdot\alpha\cdot \mathrm{tr}(\nabla_{\bm\Theta} f(\bm\Theta_0)^H\mathbf D_0),
\end{equation}
where $ \zeta\in(0,0.5) $ is a constant shrinkage factor. Substituting  (\ref{grad}) into \eqref{armijo}, we have
\begin{equation}\label{key}
	\alpha^2\big\|\mathbf H^H\mathbf H\mathbf G^H\mathbf E^H\mathbf E\big\|_F^2+\alpha\cdot(2-\zeta)\big\|\mathbf H\mathbf G^H\mathbf E^H\big\|_F^2\geq 0,
\end{equation}
which always holds when $ \alpha\geq 0 $. To ensure the solution lies at the boundary of convex set $ \mathcal S $, we set $\alpha=\frac{\sqrt{N}}{\|\mathbf H\mathbf G^H\mathbf E^H\|_F} $ and we end up with our proposed low-complexity solution of \eqref{P3} as follows:
\vspace{-0.3cm}
\begin{equation}\label{hsolution}
\bm\Theta=	\frac{\sqrt{N}}{\|\mathbf H\mathbf G^H\mathbf E^H\|_F} \mathbf H\mathbf G^H\mathbf E^H.
\vspace{-0.1cm}
\end{equation}
However, the relaxed solutions do not belong to any feasible constraint set $ \mathcal M_i,\forall i\in\{1,2,3\} $. In the next subsection, we will propose a method to project \eqref{hsolution} onto the constraint set $ \mathcal M_1$, satisfying the symmetric unitary constraints. Then, we will extend the solution to the cases of  $ \mathcal M_2$ and $ \mathcal M_3$.
\vspace{-0.2cm}
\subsection{Symmetric Unitary Projection}   

After obtaining the feasible solution of problem \eqref{P3}, we can then construct a feasible  solution for problem \eqref{P2} by projecting $ \mathbf \Theta $ onto $ \mathcal M_1$. To achieve this, we introduce the definition of symmetric projection for a square matrix $ \mathbf Z_1 $, given as 
	\vspace{-0.2cm}
\begin{equation}\label{key}
	 \mathrm{sym}(\mathbf Z_1)\triangleq\frac{1}{2}(\mathbf Z_1+\mathbf Z_1^T)=\arg\min_{\mathbf Q=\mathbf Q^T}\|\mathbf Z_1-\mathbf Q\|_F^2, 
	 \vspace{-0.2cm}
\end{equation}
Additionally, for a square matrix $ \mathbf Z_2 $, we introduce the unitary projection \cite{Manton2002}, defined as
\vspace{-0.1cm}
\begin{equation}\label{key}
	\mathrm{uni}(\mathbf Z_2)\triangleq\mathbf U\mathbf V^H=\arg\min_{\mathbf Q\mathbf Q^H=\mathbf I}\|\mathbf Z_2-\mathbf Q\|_F^2.
	\vspace{-0.1cm}
\end{equation}
where $ \mathbf U,\mathbf V $ are unitary matrices that satisfy the singular value decomposition (SVD) of $ \mathbf Z_2 $, i.e., $ \mathbf Z_2=\mathbf U \mathbf S\mathbf V^H $, and $ \mathbf S $ is a diagonal matrix. 

Suppose that $ R$ is the rank of matrix $ \mathrm{sym}(\mathbf Z)$ and $ \mathrm{sym}(\mathbf Z)= \mathbf U \mathbf S\mathbf V^H$, we are able to partition the matrices $ \mathbf U $ and $ \mathbf V $ as $ \mathbf U=[\mathbf U_R,\mathbf U_{N-R}] $ and $ \mathbf V=[\mathbf V_R,\mathbf V_{N-R}] $, respectively. Then, the symmetric unitary projection operation is defined as
\vspace{-0.1cm}
\begin{equation}\label{symuni}
	\mathrm{symuni}(\mathbf Z)=\mathrm{uni}(\mathrm{sym}(\mathbf Z))\triangleq\widehat{\mathbf U}\mathbf V^H,
	\vspace{-0.1cm}
\end{equation}
where $ \widehat{\mathbf U}\triangleq[\mathbf U_R,\mathbf V_{N-R}^*] $. Note that if $ \mathrm{sym}(\mathbf Z) $ is a rank-deficient matrix, then $ \widehat{\mathbf U}\mathbf V^H $ represents one of the unitary projections for $ \mathrm{sym}(\mathbf Z) $. Note that the proposed symmetric unitary projection is the closest point projection as shown in the following Proposition \ref{pro sym} .
\begin{proposition}\label{pro sym}
	Let $ \mathbf Z\in\mathbb C^{N\times N} $ be a square matrix, we have 
	\vspace{-0.1cm}
	\begin{equation}\label{pro clo}
	\mathrm{symuni}(\mathbf Z)=\arg\min_{\mathbf Q\in\mathcal M_1}\|\mathbf Z-\mathbf Q\|_F^2.
	\vspace{-0.5cm}
	\end{equation}
\end{proposition}
\textit{Proof:} See Appendix \ref{Proof of P1}. 

Using the symmetric unitary projection \eqref{symuni}, we immediately derive the following Lemma \ref{con}.
\begin{lemma}\label{con}
	For any non-zero constant $ \rho $ and matrix $ \mathbf Z $, the symmetric unitary projection of $ \mathbf Z $ satisfies
	\begin{equation}\label{key}
		\mathrm{symuni}(\rho \mathbf Z)=\widehat{\mathbf U}\mathbf V^H=\mathrm{symuni}(\mathbf Z).
	\end{equation}
\end{lemma}

Based on Lemma \ref{con} and Proposition \ref{pro sym}, we derive a closed form passive beamforming solution for problem \eqref{P2} when $\bm\Theta\in\mathcal M_1$, which is given by
\vspace{-0.2cm}
\begin{equation}\label{RIS}
	\mathbf \Theta=\mathrm{symuni}(\mathbf H\mathbf G^H\mathbf E^H).
	\vspace{-0.2cm}
\end{equation}

\subsection{Extension to Different Types of BD-RIS}

Next, we extend the solution \eqref{RIS} to solve problem \eqref{P2} when $\bm\Theta\in\mathcal M_2$ or $\bm\Theta\in\mathcal M_3$.

\textit{Group connected Passive Beamforming}: To project $ \eqref{hsolution} $ on to $\mathcal M_3$, we first block-diagonalize it and then apply symmetric unitary projection in each block. The block-diagonalization operation for any given square matrix $ \mathbf Z $ is defined as
\vspace{-0.1cm} \begin{equation}\label{key}
		\mathrm{blkdiag}(\mathbf Z)
		=\mathrm{diag}\{\mathbf 1,\cdots,\mathbf 1\}\odot\mathbf Z,
		\vspace{-0.1cm}
\end{equation}
where $ \mathbf 1 $ is a $ N_g $-dimensional square matrix filled with $ 1 $ and $ \odot $ denotes the Hadamard product. Let $ \mathrm{diag}\{\mathbf X_1,\cdots,\mathbf X_G\}=\mathrm{blkdiag}(\mathbf H\mathbf G^H\mathbf E^H ) $, then the solution of passive beamforming matrix in \eqref{P2} satisfying constraint set $ \mathcal M_3 $ is given by
\begin{equation}\label{group}
	\mathbf \Theta=\mathrm{diag}\{\mathrm{symuni}(\mathbf X_1),\cdots, \mathrm{symuni}(\mathbf X_G) \}.
\end{equation}  

\textit{Single connected Passive Beamforming}:  The single connected RIS should satisfy $\bm\Theta\in\mathcal M_2$, which can be considered as a special case of the group connected RIS when the number of user group is equal to the number of RIS elements, i.e., $ G=N $. Therefore, \eqref{group} reduces to the solution for constraint set $ \mathcal M_2 $ when $ N_g=1, \forall g\in\{1,\cdots,G\} $ in each group, which is given by
\vspace{-0.1cm}
\begin{equation}\label{diag}
	\mathbf \Theta=\mathrm{diag}\{e^{\mathrm{1j}\angle x_1},\cdots,e^{\mathrm{1j}\angle x_N}\},
\end{equation}
where $ \mathbf{X}_n,\forall  n\in\{1,\cdots,N\} $ reduces from matrix to scalar $ x_n $, $ \mathrm{1j} $ denotes the imaginary unit, and $ \angle x_n$ returns the phase angle for a complex number $ x_n. $ 

\section{The Proposed Two-Stage Beamforming Design Framework}
In this section, we apply the proposed low-complexity passive beamforming approach to tackle the following joint active and passive beamforming optimization problem for sum-rate maximization:
\begin{subequations}\label{P1}
	\begin{align}
		\max_{\bm\Theta,\mathbf W}\,\, &\sum_{k=1}^K \log\left(1+\frac{|\mathbf f_k^H\w_k |^2}{\sum_{j=1,j\neq k}^K |\mathbf f_k^H\w_j |^2+\sigma_k^2}\right)\\
		\text{s.t.}\,\,&\bm\Theta\in\mathcal M_i, \forall i\in\{1,2,3\},\\	&\|\mathbf W\|_F^2\leq P_t,
	\end{align}
\end{subequations}
where $ \mathbf f_k^H\triangleq\mathbf g^H_k+\mathbf h_k^H\bm\Theta\E  $ is the effective channel between the BS and user $ k $.
Problem \eqref{P1} is a challenging optimization problem due to its non-convex objective function and the non-convex constraints $ \mathcal M_i,\forall i\in\{1,2,3\} $. Existing works typically adopt an AO framework, which iteratively updates the active and passive beamforming matrices until convergence. 

\par 
In contrast to the conventional AO framework, we adopt a different approach to solve \eqref{P1} by decomposing the joint active and passive beamforming design process into two distinct stages. In the first stage, we focus on the passive beamforming design with the goal of enhancing the wireless communication environment by solving problem (2). Once the passive beamforming is obtained, we move on to the second stage, where we optimize the active beamforming to maximize the system sum-rate. The details of the proposed two-stage beamforming design framework are provided below:
\begin{itemize}
	\item \textit{Stage 1:} Design the passive beamforming $ \bm\Theta $ based on  \eqref{RIS} (fully connected), \eqref{group} (group connected), and \eqref{diag} (single connected);

\item\textit{Stage 2:} With the designed $ \bm\Theta $, solve the active beamforming design problem by classical beamforming algorithms.
\end{itemize}
 The subproblem in \textit{stage 2} related to the active beamforming matrix $ \mathbf W $ is a conventional beamforming optimization problem. Numerous algorithms exist to obtain a locally optimal solution for this problem, such as the fractional programming (FP) approach \cite{Shen2018}.
For a  more practical design, classical low-complexity beamforming algorithms could also be used, such as the regularized zero forcing (RZF) beamforming $ \mathbf W_{\text{RZF}}=\mathbf F(\mathbf F^H\mathbf F+\eta\mathbf I)^{-1} $, where $ \eta $ is a regularization parameter \cite{oestges2010mimo}.

The complexity of the scattering matrix design in the fully connected architecture is primarily influenced by SVD of the symmetric unitary projection, which has a complexity order of $ \mathcal{O}(N^3) $. For the group connected architecture, the complexity order is $ \mathcal{O}(NN_g^2) $, and for the single connected architecture, it is $ \mathcal{O}(N) $. Regarding the active beamfoming matrix optimization, the FP algorithm has a complexity of $ \mathcal{O}(I_\mu I_F KL^3) $ \cite{Shen2018}, where $ I_\mu $ and $ I_F $ denote the iteration numbers for a bisection search to find the optimal dual variable and the updating loop for the FP approach, respectively. Furthermore, the complexity reduces to $ \mathcal{O}( K^2L) $ when we replace FP with the RZF algorithm.


\section{Numerical Results}
 
In this section, we assess the performance of the proposed design. The channels are characterized by both small-scale fading, assumed to follow Rayleigh fading, and large-scale fading, modeled as the distance-dependent path loss model. The path loss model is represented as $ \zeta(d) = \zeta_0 d^{-\gamma} $, where $ \zeta_0 = -30 $ dB is the reference path loss at a distance of 1 meter, $ d $ is the link distance, and $ \gamma $ is the path loss exponent. Specifically, the path loss exponents for the BS-to-user, BS-to-RIS, and RIS-to-user channels are set to 3.5, 2, and 2.2, respectively. The distances for these three channels are fixed at 150, $ 50\sqrt{2} $, and $ 50\sqrt{5} $ meters. A convergence tolerance of $ 10^{-3} $ is employed for the baseline iterative algorithms, and the noise power is set to $ \sigma_k^2 = -80 $ dBm for all users $ k \in \mathcal{K} $. We consider $ L=K=4 $ and a transmit SNR of $ 20 $ dB. To ensure reliable results, the simulation outcomes are averaged over 100 random channel realizations. 

Problem \eqref{P2} can be solved using both the projections of \eqref{optimal} and \eqref{hsolution}. The former is the optimal solution of \eqref{P3} while the latter is the low-complexity solution proposed in this work. The projections of \eqref{hsolution} are labeled as \textbf{PoP} and those of \eqref{optimal} are labeled as \textbf{PoO}. We also evaluate the algorithm proposed in \cite{Nerini2023a}, which \textbf{directly designs} the passive beamforming matrix satisfying $ \mathcal M_1 $ using an AO framework. This algorithm is also considered as a baseline scheme and is denoted by \textbf{DD}. Moreover, we use \textbf{FC/SC/GC} to denote the fully/single/group connected arcgitectures of BD-RIS, which corresponds to the solutions of problem \eqref{P2} with constraints $ \mathcal M_i,\forall i\in\{1,2,3\} $, respectively.

We employ the proposed two-stage beamforming framework for solving problem \eqref{P1}, denoting it as \textbf{proposed 1} with FP and \textbf{proposed 2} with RZF. The algorithms in \cite{Li2023} and \cite{Fang2022} are referred to as \textbf{baseline 1} and \textbf{baseline 2}, respectively. Both baselines, following the AO framework, differ in their passive beamforming subproblem-solving approaches, with \textbf{baseline 1} using manifold optimization and \textbf{baseline 2} using quasi-Newton method.


\begin{figure}[t!]
	\centering
	\vspace{-0.7cm}
	\includegraphics[width=7.9cm]{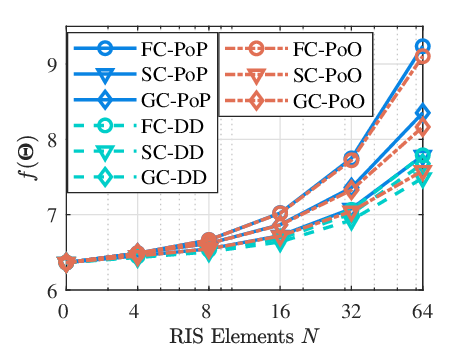}
	\caption{The sum of effective channel gains among users versus the number of RIS elements $ N $.}
	\label{fig1}
	\vspace{-0.3cm}
\end{figure}

Fig. \ref{fig1} illustrates the performance of the proposed closed-form passive beamforming design for BD-RIS. It is evident from the figure that our proposed method consistently outperforms the baselines from \cite{Nerini2023a} for all three BD-RIS architectures, regardless of the number of RIS elements.  We also observe that the projections based on \eqref{hsolution} achieve almost the same performance compared to those based on \eqref{optimal} when $ N\leq 16 $ and they can achieve better performance than those of \eqref{optimal} when $ N\geq32 $. The results demonstrate the effectiveness of our proposed low-complexity algorithm for solving problem 
\eqref{P2}. Note the the projections of the optimal solution for problem \eqref{P3} may not be the optimal solutions for problem \eqref{P2} due to the highly non-convexity of the considered constraints sets $\mathcal M_i, \forall i\in\{1,2,3\} $.

\begin{figure}[t!]
	\centering
	\vspace{-0.2cm}
	\includegraphics[width=7.9cm]{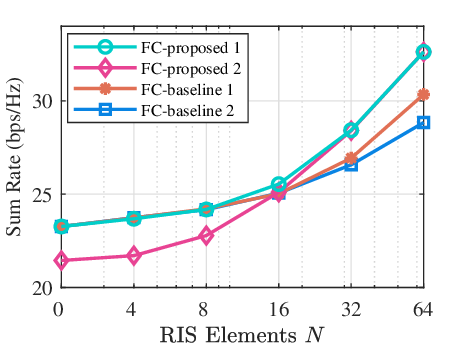}
	\caption{Sum-rate versus RIS elements $ N $ under the fully connected architecture when $ L=K=4 $ and transmit SNR is $ 20 $ dB. }
	\label{fig2}
\end{figure}
\begin{figure}[t!]
	\centering
	\vspace{-0.2cm}
	\includegraphics[width=7.9cm]{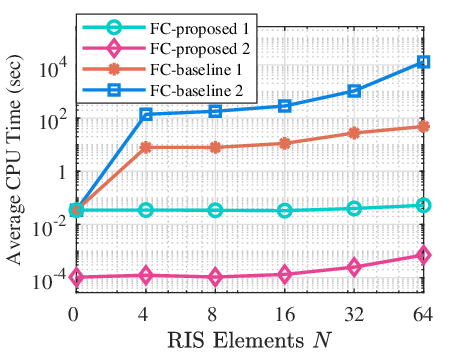}
	\caption{Average CPU time versus RIS elements $ N $ under the fully connected architecture when $ L=K=4 $ and transmit SNR is $ 20 $ dB.}
	\label{fig3}
\end{figure}
 
Fig. \ref{fig2} and Fig. \ref{fig3} respectively show the performance of different strategies in terms of sum-rate and average CPU time, with respect to the number of passive elements at the fully connected BD-RIS. The proposed two-stage algorithm, based on FP (\textbf{proposed 1}), achieves the same or superior sum-rate compared to the two baselines while significantly reducing computation time. Intriguingly, when using a large number of BD-RIS elements, i.e., $ N\geq 16 $, the combination of the proposed solution \eqref{RIS} and the RZF precoding (\textbf{proposed 2}) attains a comparable sum-rate to the AO-based methods but with notably shorter computation times. The proposed two-stage beamforming framework places more emphasis on the role of BD-RIS in shaping user channels, rather than treating it solely as an additional optimization variable for the sum-rate maximization problem. The simulation results further demonstrate the substantial benefits of our proposed two-stage framework.

\section{Conclusion}
  This study introduces a closed-form low-complexity algorithm to design passive beamforming matrix for BD-RIS. The solution is also extended to design a two-stage algorithm for active and passive beamforming design. Numerical results show that the proposed two-stage beamforming framework significantly reduces computational complexity while achieving a higher sum-rate compared to existing AO-based algorithms. The proposed framework holds potential for practical implementation in BD-RIS aided communication systems, offering improved efficiency and performance gain.

\begin{appendices}

	
\section{Proof of Proposition \ref{pro sym}}\label{Proof of P1} 
Let $ \mathbf X=\mathrm{sym}(\mathbf Z)=\frac{1}{2}(\mathbf Z+\mathbf Z^T)=	\mathbf U  \mathbf S\mathbf V^H $. The right-hand side of \eqref{pro clo} can be rewritten as
\begin{equation*}\label{key}
	\begin{split}
		&\arg\min_{\mathbf Q\in\mathcal M_1} \|\mathbf Z-\mathbf X+\mathbf X-\mathbf Q\|_F^2\\
		&\mathop=\limits^{(a)}\arg\min_{\mathbf Q\in\mathcal M_1}\|\mathbf X-\mathbf Q\|_F^2+2\Re\{\mathrm{Tr}((\mathbf Z-\mathbf X)(\mathbf X-\mathbf Q)^H)\}\\
		&\mathop=\limits^{(b)}\arg\min_{\mathbf Q\in\mathcal M_1} \|\mathbf X-\mathbf Q\|_F^2.	
	\end{split}
\end{equation*}
Step $ (b) $ follows the fact that the inner product of the skew-symmetric matrix $ \mathbf Z-\mathbf X $ and symmetric matrix $ \mathbf X- \mathbf Q  $ is zero, i.e., $ \mathrm{Tr}((\mathbf Z-\mathbf X)(\mathbf X-\mathbf Q)^H)=0 $. Note that $ \mathcal M_1 $ is a subset of $ \{\mathbf \Theta|\mathbf \Theta\mathbf \Theta^H=\mathbf I\} $, $ \widehat{\mathbf U}\mathbf V^H=\arg\min_{\mathbf Q\mathbf Q^H=\mathbf I}\|\mathbf X-\mathbf Q\|_F^2$ can therefore reach a lower bound of $ \arg\min_{\mathbf Q\in\mathcal M_1} \|\mathbf X-\mathbf Q\|_F^2	 $. This lower bound is tight since $\widehat{\mathbf U}\mathbf V^H\in\mathcal M_1  $.

Substituting the SVD of $ \mathbf X $ into $\mathbf X=\mathbf X^T $, we have 
\begin{equation}\label{svd}
	\mathbf U  \mathbf S\mathbf V^H= \mathbf V^*\mathbf S\mathbf U^T.
\end{equation}
Left-multiplying and right-multiplying \eqref{svd} by $ \mathbf U^H\! $ and $ \mathbf U^* $ respectively yields $ \mathbf S\mathbf V^H\mathbf U^*\!=\!\mathbf U^H\mathbf V^*\mathbf S $, which can be blocked as
\begin{equation}\label{key}
	\begin{bmatrix}
		\mathbf S_R\mathbf V_R^H\mathbf U_R^*& \mathbf S_R\mathbf V_R^H\mathbf U_{N-R}^*\\
		\mathbf 0&\mathbf 0
	\end{bmatrix}=\begin{bmatrix}
	\mathbf U_R^H\mathbf V_R^*\mathbf S_R&\mathbf 0\\
	\mathbf U_{N-R}^H\mathbf V_R^*\mathbf R&\mathbf 0
\end{bmatrix}.
\end{equation}
Define $ \mathbf Y_R=\mathbf U_R^H\mathbf V_R^* $, its non-diagonal entry $ y_{i,j},i\neq j,\forall i,j\in\{1,\cdots,R\} $ must satisfies $  y_{i,j}=\frac{s_i}{s_j} y_{j,i} $.
 Similarly, left-multiplying and right-multiplying \eqref{svd} by $ \mathbf V^T $ and $ \mathbf V $ respectively yields $ \mathbf V^T\mathbf U\mathbf S=\mathbf S\mathbf U^T\mathbf V $, we have
\begin{equation}\label{a2}
\begin{bmatrix}
		\mathbf V_R^T\mathbf U_R\mathbf S_R&\mathbf 0\\
		\mathbf V_{N-R}^T\mathbf U_R\mathbf R&\mathbf 0
	\end{bmatrix}=\begin{bmatrix}
	\mathbf S_R\mathbf U_R^T\mathbf V_R& \mathbf S_R\mathbf U_R^T\mathbf V_{N-R}\\
	\mathbf 0&\mathbf 0
\end{bmatrix}.
\end{equation}
\eqref{a2} leads to $ \mathbf U_R^H\mathbf V_{N-R}^*=\mathbf 0 $ and $ y_{j,i}^*=\frac{s_i}{s_j} y_{i,j}^* $. Combining with $  y_{i,j}=\frac{s_i}{s_j} y_{j,i} $, this implies that $ y_{i,j}=0,i\neq j $ and $ \mathbf Y_R $ is a diagonal matrix. Recall that $ \widehat {\mathbf U}=[\mathbf U_R,\mathbf V_{N-R}^*] $, it is easy to verify that $ \widehat {\mathbf U} $ is a unitary matrix and $ \widehat {\mathbf U}^H\mathbf V^* $ is a diagonal matrix, i.e., $\widehat {\mathbf U}^H\mathbf V^*= \mathbf V^H\widehat {\mathbf U}^* $.

Now the symmetric unitary projection of $ \mathbf Z $ is able to be written as $ 	\mathrm{symuni}(\mathbf Z)=\widehat {\mathbf U}\mathbf V^H=\mathbf V^*\widehat {\mathbf U}^T=\mathrm{symuni}(\mathbf Z)^T $. Together with $ \mathrm{symuni}(\mathbf Z)	\mathrm{symuni}(\mathbf Z)^H=\widehat {\mathbf U}\mathbf V^H\mathbf V\widehat {\mathbf U}^H=\mathbf I $, we complete this proof.$ \hfill \blacksquare $
\end{appendices}

\bibliographystyle{IEEEtran}
\bibliography{reference}
%
%

\end{document}